\documentclass[10pt]{allhands}

\PassOptionsToPackage{hyphens}{url}
\usepackage{microtype}
\usepackage{graphicx}
\usepackage{booktabs} %
\usepackage{xcolor} %
\usepackage{colortbl} %
\usepackage{color-edits}

\usepackage{tikz}
 \usetikzlibrary{positioning,fit,calc,arrows.meta,backgrounds}

\usepackage[cachedir=.]{minted}
\setminted{
  fontsize=\footnotesize,
  breaklines=true,
  frame=single,
  xleftmargin=0.5em,
  bgcolor=gray!5,
  linenos=false,
  python3=true
}
\usepackage{listings}
\addauthor{vc}{blue}
\addauthor{gn}{magenta}

\definecolor{darkgreen}{RGB}{0,128,0}

\usepackage{fontawesome5}
\usepackage{amsfonts}
\usepackage{amsmath}
\usepackage{amsthm}
\usepackage{wrapfig}

\theoremstyle{plain}

\theoremstyle{definition}

\theoremstyle{remark}

\usepackage{multirow}
\usepackage{tabularx}
\usepackage{enumitem}
\usepackage{xspace}
\usepackage[most]{tcolorbox}
\usepackage{subcaption}

\usepackage[hyphens]{url}

\usepackage{inconsolata}
\usepackage{makecell}
\usepackage{array}

\tcbset{
  lessonstyle/.style={
    colback=gray!5!white,
    colframe=gray!50!black,
    boxrule=0.5pt,
    arc=2pt,
    left=6pt,right=6pt,top=4pt,bottom=4pt,
    fonttitle=\bfseries,
    title={#1}
  }
}

\newenvironment{itemize*}%
 {\leftmargini=10pt\begin{itemize}%
  \setlength{\itemsep}{0pt}%
  \setlength{\parskip}{0pt}%
  }%
 {\end{itemize}}
\newenvironment{enumerate*}%
 {\begin{enumerate}%
  \setlength{\itemsep}{0pt}%
  \setlength{\parskip}{0pt}}%
 {\end{enumerate}}

\newcommand{\cmark}{\textcolor{darkgreen}{\textbf{$\checkmark$}}}
\newcommand{\simmark}{\textcolor{orange}{$\sim$}}
\newcommand{\xmark}{\textcolor{red}{$\times$}}

\newcommand{\sref}[1]{\S\ref{#1}}
\newcommand{\fref}[1]{Fig.~\ref{#1}}
\newcommand{\tref}[1]{Tab.~\ref{#1}}

\tcbset{
  principlebox/.style={
    colback=gray!4!white,
    colframe=gray!40!black,
    boxrule=0.4pt,
    arc=2pt,
    left=6pt,right=6pt,top=4pt,bottom=4pt,
    title={V1 Design Principle}
  }
}

\begin{document}

\title{The OpenHands Software Agent SDK: A Composable and Extensible Foundation for Production Agents}

\author[]{Xingyao Wang}
\author[]{Simon Rosenberg}
\author[]{Juan Michelini}
\author[]{Calvin Smith}
\author[]{Hoang Tran}
\author[]{Engel Nyst}
\author[]{Rohit Malhotra}
\author[]{Xuhui Zhou}
\author[]{Valerie Chen}
\author[]{Robert Brennan}
\author[]{Graham Neubig}

\contact{\{xingyao, graham\}@openhands.dev}

\abstract{%
Agents are now used widely in the process of software development, but building production-ready software engineering agents is a complex task.
Deploying software agents effectively requires flexibility in implementation and experimentation, reliable and secure execution, and interfaces for users to interact with agents.
In this paper, we present the \textbf{OpenHands Software Agent SDK}, a toolkit for implementing software development agents that satisfy these desiderata.
This toolkit is a complete architectural redesign of the agent components of the popular \textbf{OpenHands} framework for software development agents.

To achieve flexibility, we design a \emph{simple interface for implementing agents} that requires only a few lines of code in the default case, but is easily extensible to more complex full-featured agents with features such as custom tools, memory management, and more.
For security and reliability, it delivers \emph{seamless local-to-remote execution portability}, integrated REST/WebSocket services.
For interaction with human users, it can \emph{connect directly to a variety of interfaces}, such as visual workspaces (VS Code, VNC, browser), command-line interfaces, and APIs. 
Compared with existing SDKs from OpenAI, Claude and Google, OpenHands uniquely integrates native sandboxed execution, lifecycle control, model-agnostic multi-LLM routing, and built-in security analysis. 
We validate the architecture empirically: production deployment data shows that V1 substantially reduces system-attributable failures over V0 with negligible event-sourcing overhead, and evaluations across multiple models and benchmarks demonstrate strong agent performance.
Put together, these elements allow the OpenHands Software Agent SDK to provide a practical foundation for prototyping, unlocking new classes of custom applications, \emph{and} reliably deploying agents at scale.

}

\metadata[Software Agent SDK]{\url{https://github.com/OpenHands/software-agent-sdk}}
\metadata[Benchmarks]{\url{https://github.com/OpenHands/benchmarks}}

\accepted{MLSys 2026}
\newcommand{\acmbadgeurl}{https://www.acm.org/publications/policies/artifact-review-and-badging-current}
\badge[\acmbadgeurl]{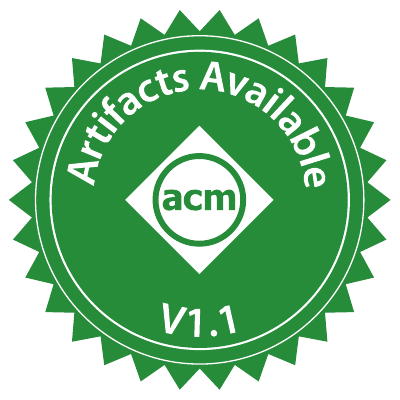}
\badge[\acmbadgeurl]{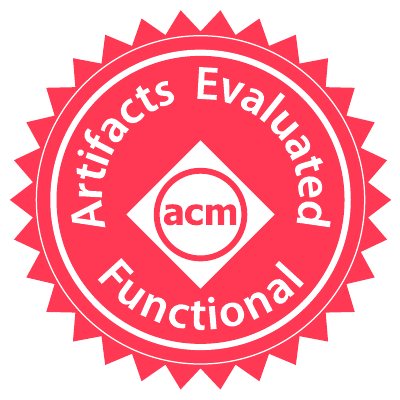}
\badge[\acmbadgeurl]{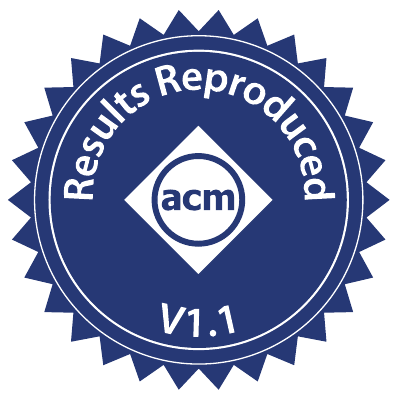}

\maketitle

\section{Introduction}
\label{sec:introduction}
\begin{figure*}[!bht]
\centering

\begin{subfigure}[b]{0.49\textwidth}
  \centering
  \resizebox{\textwidth}{!}{\includegraphics[trim=3cm 4cm 3cm 4cm, clip]{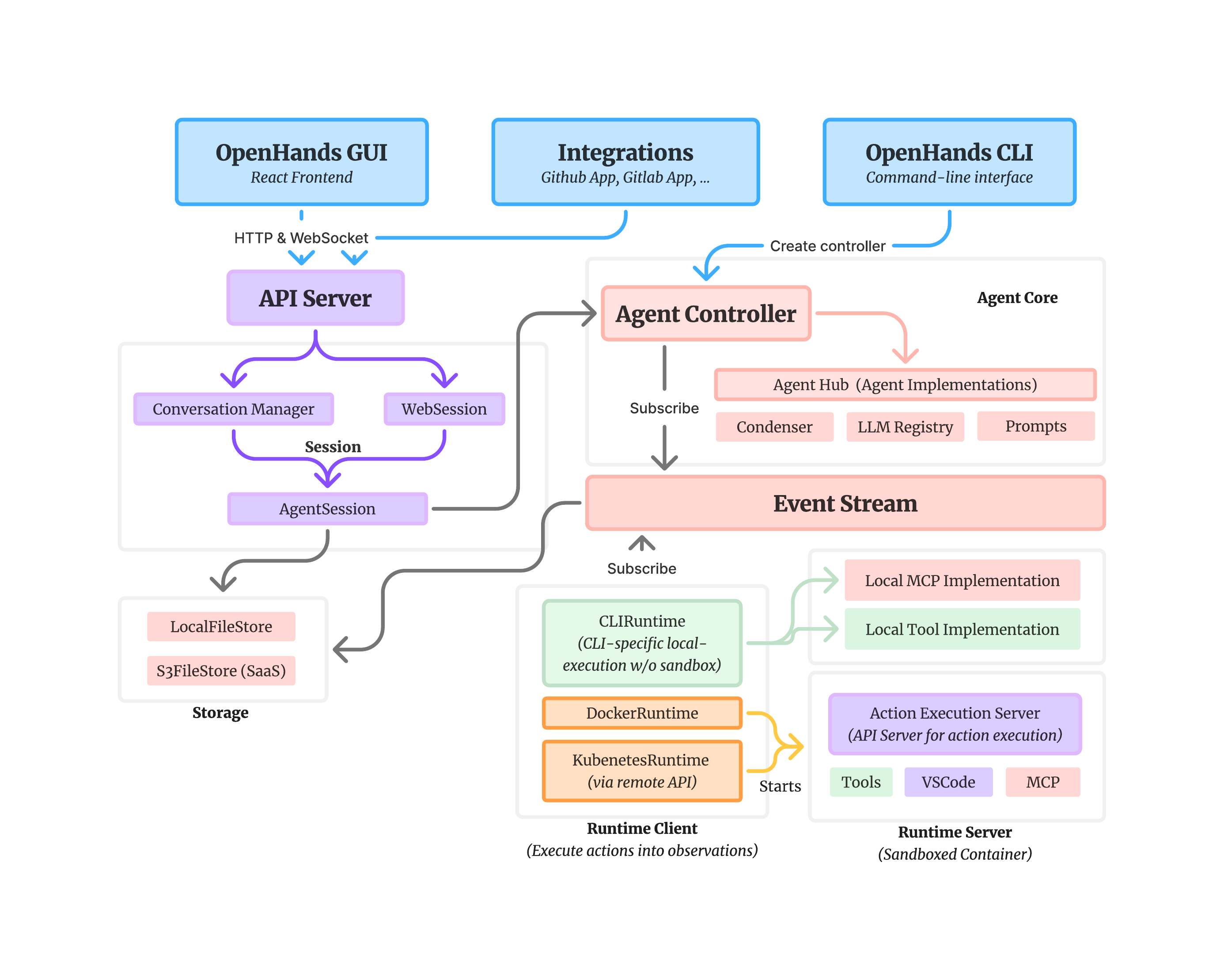}}
  \caption{OpenHands V0: Monolithic architecture with tightly coupled components and a mandatory sandboxing model. The design assumed all executions occur in a sandbox, making later support for local execution workflows (CLI) cumbersome — requiring special-case handling in the CLI runtime and duplicated local implementations of MCP and tools.}
  \label{fig:v0-architecture}
\end{subfigure}
\hfill
\begin{subfigure}[b]{0.49\textwidth}
  \centering
  \resizebox{\textwidth}{!}{
    \includegraphics[trim=3cm 4cm 3cm 4cm, clip]{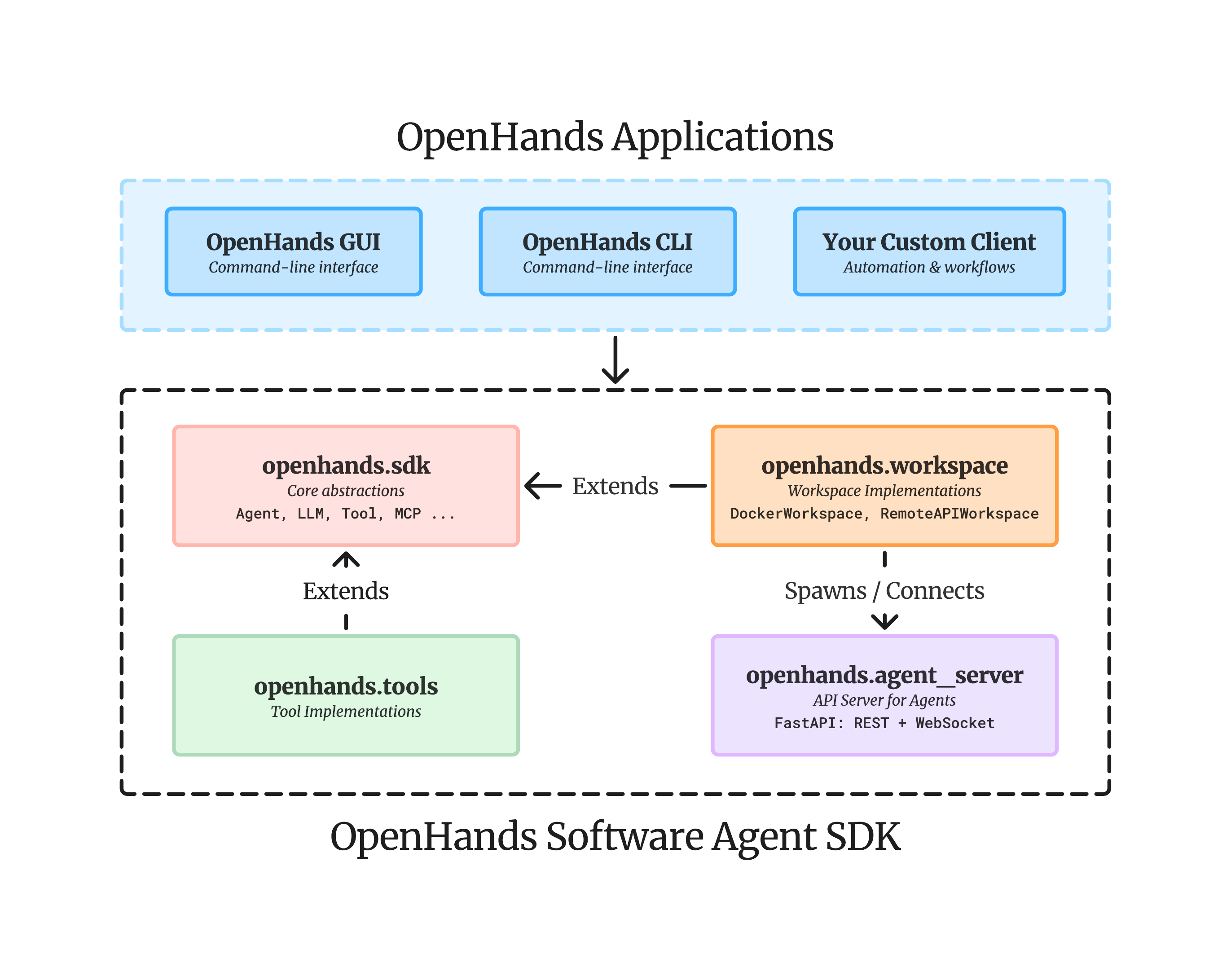}
  }
  \caption{OpenHands V1: Modular SDK architecture with four decoupled packages. Applications consume software agent SDK; \texttt{tool} and \texttt{workspace} packages extend \texttt{sdk}'s abstraction with actual implementations; the \texttt{workspace} package spawns and connects to \texttt{agent server}.}
  \label{fig:v1-architecture}
  \label{fig:sdk-ecosystem}
\end{subfigure}

\caption{Architecture Evolution from OpenHands V0 to V1. V0’s monolithic, sandbox-centric design tightly coupled components and required duplicated local implementations. V1 refactors this into a modular SDK with clear boundaries, opt-in sandboxing, and reusable agent, tool, and workspace packages.
The color of each component in V0 roughly corresponds to its modular counterpart in V1.
}
\label{fig:v0-v1-comparison}
\end{figure*}

In software engineering, AI agents have evolved from assistive tools~\citep{copilot2021github,cursor2024} to autonomous systems capable of hours-long async execution on complex tasks (e.g., Devin~\citep{devin2024cognition}, Claude Code~\citep{claudecode}, OpenHands~\citep{wang2025openhands}).
Reliably deploying autonomous agents in production requires \textit{system foundations}---including durable state management, safe execution in a sandbox, and consistent behavior across environments ranging from local to containerized cloud deployments---that were unnecessary for earlier assistive tools that rely on largely local user-driven workflows.
Yet no consensus exists on how to architect these foundations: existing SDKs from major providers differ widely in isolation models, state management, and deployment assumptions, leaving practitioners without a well-validated reference design.

OpenHands~\citep{wang2025openhands} demonstrated that open-source software agents can achieve broad adoption and contributions, reaching over 64k GitHub stars and hundreds of contributors in just 18 months~\citep{neubig2025-one-year-of}.
As the project scaled, the original monolithic architecture---what we refer to as OpenHands V0---exposed growing architectural tensions.
The early design was driven by the need for fast prototyping and iteration, combining agent logic, evaluation, and applications in a single codebase. 
Over time, this approach led to rigid sandboxing assumptions, sprawling mutable configuration, and tight coupling between research and production, eventually making a full architectural redesign unavoidable.

Guided by these lessons, \textbf{OpenHands~V1} introduces a new architecture grounded in four design principles that directly address the limitations of V0:
\begin{itemize}[itemsep=2pt,topsep=0pt,parsep=0pt,partopsep=0pt,leftmargin=8pt]
    \item \textbf{Optional isolation.} The agent runs locally by default but can switch to a sandboxed environment when additional safety or resource control is required. 
    \item \textbf{Stateless by default, one source of truth for state.} All components---agents, tools, LLMs, etc---are immutable and validated at construction, while a single conversation state object records all mutable context. This design ensures reliable recovery of agent sessions.
    \item \textbf{Strict separation of concerns.} The agent core is decoupled from applications so that downstream systems (CLI, Web UI, GitHub App) use it as a shared library rather than duplicating logic.
    \item \textbf{Two-layer composability.} Developers can compose independent deployment packages (SDK, Tools, Workspace, Server) and extend the SDK safely by adding or replacing typed components such as tools or agents.
\end{itemize}

Building on these principles, we introduce \textbf{OpenHands~V1}, a complete software-agent ecosystem---including CLI and GUI applications---built on a shared foundation: the \textbf{OpenHands Software Agent SDK} (\fref{fig:sdk-ecosystem}).
We present the SDK as a \emph{reference architecture for production agent systems}, codifying lessons from 18 months of open-source development and production deployment into an empirically validated design.
The SDK defines an \textit{event-sourced state model} with deterministic replay, an \textit{immutable configuration} for agents, and a \textit{typed tool system} with MCP integration. Its \textit{workspace abstraction} enables the same agent to run locally for prototyping or remotely in secure, containerized environments with minimal code changes.
Unlike prior library-only SDKs \citep{anthropic-agents,openai_agents_2024}, OpenHands includes a \textit{built-in REST/WebSocket server} for remote execution and a \textit{suite of interactive workspace interfaces} for human inspection and control.

\begin{figure}[!htb]
\begin{minted}[fontsize=\scriptsize, breaklines, linenos]{python}
from openhands.sdk import LLM, Conversation
from openhands.tools.preset.default import get_default_agent
llm = LLM(model="openhands/claude-sonnet-4-5-20250929", api_key="...")
agent = get_default_agent(llm=llm)
conversation = Conversation(agent=agent, workspace="/path/to/project")
conversation.send_message("Write 3 facts about this project into FACTS.txt.")
conversation.run()
\end{minted}
\caption{Minimal example of OpenHands Software Agent SDK.}
\label{fig:hello-world}
\end{figure}

We systematically compare features of our SDK with those of the OpenAI Agents SDK, Claude Agent SDK, Google ADK, and LangChain (\tref{tab:sdk-comparison}), finding that our SDK uniquely combines native remote execution with sandboxing, a production server, model-agnostic multi-LLM routing across 100+ providers, a security analyzer for agent actions, and built-in QA instrumentation for production reliability. 

We support this architectural contribution with empirical evidence at two levels.
At the \emph{systems level}, a 15-day production comparison shows V1 reduces system-attributable failures by 61\% relative to V0, and event-sourcing overhead is negligible.
At the \emph{agent capability level}, evaluations across 14 language models and five benchmark categories---including SWE-Bench Verified, GAIA, SWE-Bench Multimodal, SWT-Bench, and Commit0---demonstrate that the redesigned architecture maintains strong agent performance.

The OpenHands Software Agent SDK\footnote{\url{https://github.com/OpenHands/software-agent-sdk}} and evaluation harness\footnote{\url{https://github.com/OpenHands/benchmarks}} are fully open-sourced under the MIT License.

\section{Preliminaries}

\subsection{Agent Design}

Modern AI agents can be viewed as systems that perceive, reason, and act within an environment to accomplish goals over time \citep{10.5555/1671238}. In recent work \citep{Wang2023ASO}, agent design typically centers on these interconnected components:

\begin{itemize}[noitemsep,topsep=0pt,parsep=0pt,partopsep=0pt,leftmargin=8pt]
\item \textbf{Environment and Observations.}
The environment defines the external context in which the agent operates, providing structured interfaces (e.g., APIs, files, user interfaces) from which it receives observations. These percepts—such as text, logs, or visual data—inform the agent’s situational awareness and ground its decisions in the evolving task state.

\item\textbf{State and Memory.}
An agent’s state encodes what it currently knows about the world and itself, often represented as a history of past interactions, retrieved information, and internal reasoning. 

\item\textbf{Actions and Tools.}
Actions represent how the agent influences its environment—issuing commands, executing code, or invoking APIs—each realized through the use of tools that implement these capabilities.

\end{itemize}

\subsection{Software Agents}

In software engineering contexts, these agentic components manifest concretely in systems like OpenHands~\citep{wang2025openhands}, which emulate a developer’s end-to-end workflow. 
The environment is a sandboxed workspace exposing a filesystem, terminal, and web interface, allowing the agent to observe program output and test feedback. 
Its state and memory are represented through an event log that records commands, edits, and results, providing persistent context across actions. 
The agent acts by editing files, running tests, or invoking structured tools like web browser to interact with its environment. 

\section{Challenges and Design Principles}
\label{sec:challenges}

As OpenHands evolved, it has accumulated significant architectural complexity: core agent logic, multiple CLIs, a web server for the GUI, frontend code, and diverse runtime providers (e.g., local Docker, Kubernetes in production) all coexist in a single codebase. 
Much of this architecture dated back to the project’s first weeks, before the introduction of LLM structured tool use or the Model Context Protocol (MCP; \citealt{mcp2025intro}). 
This section revisits the key architectural tensions observed in OpenHands V0 and distills them into four design principles guiding V1.

\subsection{Universal Sandboxing vs. Local Flexibility}

\textbf{V0 Challenges.}
V0 was built on the assumption that all tool calls should run inside sandboxed Docker containers for safety and reproducibility. 

However, doing so introduced multiple layers of friction as each conversation spanned two independent processes (agent and sandbox) with potentially divergent states.
Since the sandbox for tools might crash while the agent continued (or vice versa), leading to corrupted sessions. In multi-tenant deployments, resource-heavy actions from one user, such as mass webpage visits generating oversized screenshots, could exhaust container resources and crash other agents sharing the same application.

When the need arose to support local execution workflows—for example, running tools or MCP clients directly on the host machine—we had to add exceptions and bypass layers that duplicated existing logic. As shown in \fref{fig:v0-architecture}, this led to redundant local versions of MCP and tool implementations, diverging from the original sandbox-based code path.
These ad hoc extensions made the architecture increasingly brittle and poorly aligned with MCP, which assumes agents can execute actions locally with direct access to credentials, files, and IDEs.

\begin{tcolorbox}[principlebox]
\textbf{Sandboxing should be opt-in, not universal.}
V1 unifies agent and tool execution in a single process by default, aligning with MCP’s assumptions. 
When isolation is needed, the same stack can be containerized transparently. 
Sandboxing becomes opt-in—preserving flexibility without sacrificing safety.
\end{tcolorbox}

\subsection{Mutable Configuration vs. Deterministic State}

\textbf{V0 Challenges.}
V0’s configuration system mixed different domains — deployment, agent behavior, LLM routing, sandbox settings, and UI options — within overlapping layers, creating hidden dependencies and complex override logic.
The deeper issue was structural: configuration was split across multiple parallel hierarchies (CLI/headless, Web UI, GitHub App, and SaaS), each evolving its own precedence rules and assumptions. Different entry points were added incrementally, each patching configuration values in place, so two runs with identical parameters could still diverge subtly. Later attempts to unify these systems only added more override layers and inconsistencies, while the web configuration remained rigid due to its coupling with ORM models and database schemas.
These issues led to severe sprawl — 140+ fields, 15 classes, and ~2.8K lines of configuration code — a brittle system where small changes often cascaded into unrelated failures.

\begin{tcolorbox}[principlebox]
\textbf{Stateless by Default, One Source of Truth for State.}
V1 treats all agents and their components—tools, LLMs, etc—as immutable and serializable Pydantic models validated at construction. 
The only mutable entity is the \emph{conversation state}, which is a single, well-defined source of truth that tracks ongoing execution. 
This design isolates change to one place, enabling deterministic replay, strong consistency, and stable long-term recovery.
\end{tcolorbox}

\subsection{Monorepo vs. Modular SDK}

\textbf{V0 Challenges.}
OpenHands V0 was implemented as a single monolithic repository that combined the agent core, evaluation suite, and applications (frontend, backend, CLI) into one codebase. 
This monorepo simplified early development but gradually blurred boundaries between the agent and its downstream applications. 
The same repository powered the CLI, web interface, and GitHub integrations, each introducing divergent logic and configuration overrides.
Over time, the agent core absorbed application-specific branches, benchmark dependencies, and environment-specific hacks, making the system brittle and slow to evolve. 

As OpenHands gained popularity in the academic community, many benchmarks were contributed, making the agent more general but also introducing heavy dependencies and frequent version conflicts. These leaked into the main application due to mono-repo design, making deployments heavyweight and fragile.

\begin{tcolorbox}[principlebox]
\textbf{Maintain strict separation of concerns. }
V1 isolates the agent core into software engineering SDK as described in this paper.
Applications integrate via SDK APIs, allowing research to evolve independently from applications.
\end{tcolorbox}

\subsection{Monolith Logic vs. Extensible Architecture}

\textbf{V0 Challenges.}
As V0 lacked clear boundaries between the agent core and its applications, adding new behaviors in V0 often required editing the core logic or branching for specific entry points, limiting experimentation and maintainability. 
The system lacked a structured notion of composability and extensibility, forcing ad hoc hacks for even small changes.

\begin{tcolorbox}[principlebox]
\textbf{Everything should be composable and safe to extend.}
V1 makes composability a first-class design goal at two levels. 
At the \textit{deployment level}, its four modular packages—SDK, Tools, Workspace, and Agent Server—combine flexibly to support local, hosted, or containerized execution. 
At the \textit{capability level}, the SDK exposes a typed component model—tools, LLMs, contexts, etc—so developers can extend or reconfigure agents declaratively without touching the core.
\end{tcolorbox}

\section{Architecture}
\label{sec:architecture}

OpenHands V1’s architecture emerged from both the operational challenges we encountered in V0 analyzed in \sref{sec:challenges}. OpenHands V1 is a \emph{broader ecosystem} with applications like CLI and GUI. Its foundation is the \textbf{OpenHands Software Agent SDK}—a standalone developer framework comprising nine interlocking components: Event-Sourced State Management (\sref{sec:event-sourcing}), LLM (\sref{sec:llm}), Tool System (\sref{sec:tool}), Agent (\sref{sec:agent}), Context Window Management (\sref{sec:context}), Local Conversation (\sref{sec:local-conversation}), Secret Registry (\sref{sec:secret-registry}), Security and Confirmation (\sref{sec:security-confirmation}), and deployment architecture (\sref{sec:deployment}) with local and remote workspace support (\sref{sec:workspace}).
This section focuses on describing the \emph{SDK architecture} rather than the full application and explains how these components collectively support both local and production deployments.

\subsection{Modular Four-Package Design}
\label{sec:overview}

The OpenHands Software Agent SDK is organized into four Python packages that compose together based on deployment needs. 
\fref{fig:sdk-ecosystem} shows how these packages interact:
\begin{itemize}[noitemsep,topsep=0pt,parsep=0pt,partopsep=0pt,leftmargin=8pt]
  \item \texttt{openhands.sdk}: Core abstractions (Agent, Conversation, LLM, Tool, MCP, etc) and the event system.
  \item \texttt{openhands.tools}: Concrete tool implementations based on abstractions defined in \texttt{openhands.sdk}.
  \item \texttt{openhands.workspace}: Execution environments (e.g., Docker, hosted API) that extend SDK base classes.
  \item \texttt{openhands.agent\_server}: A web server exposing REST/WebSocket APIs for remote execution.
\end{itemize}

The separation addresses key production concerns that slowed down development, QA, and release in OpenHands V0: (1) \texttt{sdk} stays lightweight for diverse integration scenarios; (2) \texttt{tools} isolates slow-running tool tests from core SDK changes, speeding up development; (3) \texttt{workspace} provides optional sandboxing implementations without bloating the core; and (4) \texttt{agent\_server} offers a generic API server usable with or without containers.
This modularity enables independent testing, selective dependency management, and incremental release cycles—critical for production deployments where monolithic repositories create QA bottlenecks.

\subsection{Event-Sourced State Management}
\label{sec:event-sourcing}

At V1's core lies an event-sourcing pattern treating all interactions as immutable events appended to a log.

\textbf{Event Hierarchy.}
The event system uses a multi-level hierarchy shown in Table~\ref{tab:event-hierarchy}. At the base, \texttt{Event} provides immutable structure (ID, timestamp, source) with type-safe serialization via discriminated unions \citep{pydantic_discriminated_unions}. \texttt{LLMConvertibleEvent} adds \texttt{to\_llm\_message()} for converting events into LLM format. The action-observation loop uses \texttt{ActionEvent} for tool calls and \texttt{ObservationBaseEvent} subclasses for results -- all these are subclasses of \texttt{LLMConvertibleEvent}. Internal events (condensation, state updates, etc) inherit directly from \texttt{Event} for bookkeeping without LLM exposure.

\begin{table}[!h]
\centering
\caption{Event hierarchy organized by parent-child relationships. LLM-convertible events can be sent to the LLM, while internal events handle state management and control flow.}
\label{tab:event-hierarchy}
\resizebox{\columnwidth}{!}{%
\begin{tabular}{@{}p{6cm}p{12cm}@{}}
\toprule
\textbf{Parent Class} & \textbf{Child Classes \& Purpose} \\
\midrule
\texttt{Event} & \textit{Base:} Immutable structure, type-safe serialization \\
\midrule
\multicolumn{2}{@{}c@{}}{\textit{LLM-Convertible Events (visible to LLM)}} \\
\texttt{LLMConvertibleEvent} & \texttt{MessageEvent}: User/assistant text messages \\
  $\hookrightarrow$ \texttt{Event} & \texttt{ActionEvent}: Agent tool calls with thought \& reasoning \\
                                   & \texttt{SystemPromptEvent}: System prompt with tool schemas \\
                                   & \texttt{CondensationSummaryEvent}: Summary of forgotten events \\
                                   & \texttt{ObservationBaseEvent}: Base for tool responses \\
\midrule
\texttt{ObservationBaseEvent} & \texttt{ObservationEvent}: Successful tool execution results \\
  $\hookrightarrow$ \texttt{LLMConvertibleEvent} & \texttt{UserRejectObservation}: User rejected action \\
                                   & \texttt{AgentErrorEvent}: Agent/scaffold errors \\
\midrule
\multicolumn{2}{@{}c@{}}{\textit{Internal Events (not visible to LLM)}} \\
\texttt{Event} & \texttt{ConversationStateUpdateEvent}: State field changes \\
               & \texttt{CondensationRequest}: Trigger history compression \\
               & \texttt{Condensation}: Compression result with forgotten events \\
               & \texttt{PauseEvent}: User-requested pause \\
\bottomrule
\end{tabular}%
}
\end{table}

\textbf{ConversationState: Single Source of Truth.}
By design, components like \texttt{Agent}, \texttt{Tool}, and \texttt{LLM} are immutable and serializable—all changing variables live in \texttt{ConversationState}, making it the \textit{only} stateful component. This class maintains two types of state: (1) mutable metadata fields (e.g., \texttt{agent\_status}, \texttt{stats}, \texttt{confirmation\_policy}) stored directly in the Pydantic model, and (2) an append-only \texttt{EventLog} recording all agent interactions. A FIFO lock ensures thread-safe updates through a two-path pattern: state-only updates for metadata changes, and event-based updates that append to the log.

When persistence is configured, \texttt{ConversationState} selectively writes changes to disk. Metadata fields serialize to a single \texttt{base\_state.json} file on each modification, while \texttt{EventStore} persists events as individual JSON files to the corresponding directory. This dual-path design enables efficient incremental persistence—only new events write to disk, avoiding rewrites of large histories. Conversations resume by loading \texttt{base\_state.json} and replaying events from the directory, with agents automatically detecting incomplete conversations and continuing from the last processed event.

Empirical measurements on agent traces confirm that event-sourcing overhead is negligible relative to LLM round-trip times, with sub-millisecond persist latency and crash recovery under 20\,ms (see \sref{sec:event-sourcing-overhead} for details).

\subsection{LLM Abstraction Layer}
\label{sec:llm}
The \texttt{LLM} class provides a unified interface to language models. Through LiteLLM, it supports 100+ providers with two APIs: the standard Chat Completions API for broad compatibility and the newer OpenAI Responses API for latest reasoning models.

\textbf{Native Support for Reasoning / Extended Thinking.}
The SDK captures and processes advanced native reasoning fields from frontier models, such as \texttt{ThinkingBlock} for Anthropic's extended thinking, and \texttt{ReasoningItemModel} for OpenAI's reasoning. The SDK supports the OpenAI Responses API transparently for the agent, enabling client developers to use the agent with advanced reasoning models like GPT-5-Codex that are only available on the recently released Responses API.

\textbf{Built-in Support for Non-function-calling Models.}
For models without native function calling support, the SDK implements a \texttt{NonNativeToolCallingMixin}, which converts tool schemas to text-based prompt instructions and parses tool calls from model outputs using structured prompts and regex-based extraction. This enables models that do not support function calling to be used for agentic tasks, dramatically expanding the set of usable models.

\textbf{Multi-LLM Routing Support.}
SDK features \texttt{RouterLLM}, a subclass of \texttt{LLM} that enables the agent to use different models for different LLM requests.
Custom implementations can extend \texttt{RouterLLM} and implement \texttt{select\_llm()} to choose a different model based on different LLM inputs.
Please refer to \fref{fig:llm-router} for a pseudo-code example.

\begin{figure}[t]
\begin{minted}[fontsize=\scriptsize, breaklines, linenos]{python}
class RouterLLM(LLM):
    llms_for_routing: dict[str, LLM]  # Available models
    
    @abstractmethod
    def select_llm(self, messages: list[Message]) -> str:
        """Return key of LLM to use from llms_for_routing."""

    def completion(
        self, messages: list[Message], ... **kwargs,
    ) -> LLMResponse:
        # Select appropriate LLM
        selected_model = self.select_llm(messages)
        self.active_llm = self.llms_for_routing[selected_model]
        return self.active_llm.completion(...)

class MultimodalRouter(RouterLLM):
    def select_llm(self, messages: list[Message]) -> str:
        has_images = any(m.contains_image for m in messages)
        return "primary" if has_images else "secondary"

# Usage: route text to cheaper model, images to multimodal model
router = MultimodalRouter(llms_for_routing={
    "primary": LLM(model="claude-sonnet-4-5"),
    "secondary": LLM(model="devstral-small")})
agent = Agent(llm=router, tools=tools)
\end{minted}
\caption{Multi-LLM routing example. Routers inherit from LLM to maintain a unified interface while delegating to selected models.}
\label{fig:llm-router}
\end{figure}

\subsection{Tool System}
\label{sec:tool}

The V1 tool system provides a type-safe and extensible framework grounded in an \textbf{Action–Execution–Observation} pattern. As illustrated in \fref{fig:tool-definition}, tool usage follows a strict contract: the LLM proposes JSON tool calls, which are validated and parsed into \texttt{Action}; these are \texttt{executed} and the results are returned as \texttt{Observation}.
This abstraction unifies how the SDK supports both custom tools and MCP tools, providing a single standard interface for defining, invoking, and managing tools.

\textbf{Action–Execution–Observation Pattern.}
Each tool is defined by three well-separated components:
\begin{itemize}[itemsep=2pt,topsep=0pt,parsep=0pt,partopsep=0pt,leftmargin=8pt]
    \item \textbf{Action} — Specifies the input schema for a tool call. LLM-generated arguments are validated against a Pydantic model before execution, ensuring type safety and preventing malformed requests.
    \item \textbf{Execution} — Implements the tool’s actual logic via the \texttt{ToolExecutor}, which receives a validated \texttt{Action} and performs the underlying execution.
    \item \textbf{Observation} — Captures the output of the execution, defining a structured return schema, and converting results (or errors) into a LLM-compatible format.
\end{itemize}

\begin{figure}[!ht]
\begin{minted}[fontsize=\scriptsize, breaklines, linenos]{python}
class Schema(BaseModel):
    @classmethod
    def to_mcp_schema(cls) -> dict[str, Any]
    @classmethod
    def from_mcp_schema(cls: type[S], model_name: str, schema: dict[str, Any]) -> type["S"]

class Action(Schema):
    """Tool input with schema validation."""
    def visualize(self) -> str  # For UI display

class Observation(Schema):
    """Tool output with LLM conversion."""
    def to_llm_content(self) -> str  # For LLM context

class ToolDefinition[ActionT, ObservationT](BaseModel):
    action_type: type[Action]
    observation_type: type[Observation]
    executor: ToolExecutor
    def to_mcp_tool(self, ...) -> dict
    def to_openai_tool(self) -> dict  # ChatCompletions API format
    def to_responses_tool(self) -> dict  # Responses API format

class ToolExecutor[ActionT, ObservationT](ABC):
    """Executor function type for a Tool."""
    def __call__(self, action: ActionT) -> ObservationT
\end{minted}
\caption{Tool system structure. Actions validate inputs, executors run logic, and observations format outputs for LLMs.}
\label{fig:tool-definition}
\end{figure}

\textbf{MCP Integration.}
The same abstraction additionally enables seamless support for the MCP. MCP tools are treated as first-class SDK tools: their JSON Schemas are automatically translated into \texttt{Action} models, and their results are surfaced as structured \texttt{Observation}. 
\texttt{MCPToolDefinition} extends the standard \texttt{ToolDefinition} interface, while \texttt{MCPToolExecutor} delegates execution to FastMCP’s \texttt{MCPClient}, which manages server communication and transport details. As a result, external MCP tools behave identically to native tools—validated on input, type-safe at runtime, and serialized for LLM consumption—highlighting the flexibility and extensibility of the core tool abstraction.

\textbf{Tool Registry and Distributed Execution.}
SDK supports distributed agent architectures by decoupling \emph{tool specifications} from \emph{implementations} using a registry-based resolution mechanism. Because Python executors are non-serializable, tools are represented as lightweight \texttt{Tool} specification objects containing only a registered name and JSON-serializable parameters. Through \texttt{register\_tool(name, ToolDefinition)}, each identifier is bound to a resolver that reconstructs the full definition—including its executor—at runtime based on conversation context. This allows tool specs to \textit{cross process or network boundaries} as pure JSON, enables lazy instantiation with environment-specific state (e.g., workspace paths), and ensures a uniform interface for local and remote execution.

\subsection{Agent: Stateless Event Processor}
\label{sec:agent}

The agent abstraction separates \emph{configuration} from \emph{execution state}. Agents are defined as stateless, immutable specifications, including LLM settings, tool specifications, security policies, and agent content, that can be serialized and transmitted across process boundaries.

\textbf{Event-Driven Execution.}
Agents execute through an event-driven loop that processes conversation state step-by-step. Rather than directly returning results, agents emit structured events (e.g., messages, actions, observations) through callbacks, i.e. \texttt{on\_event(event: Event) -> None}, separating event generation from execution control. This design enables: (1) \textbf{security interleaving}—actions can be reviewed or blocked before execution based on risk analysis (\sref{sec:security-confirmation}); (2) \textbf{incremental execution}—the agent advances one step at a time, supporting pause/resume, recovery from context overflows, and condensation for long conversations; and (3) \textbf{event streaming}—intermediate results (e.g., observations and reasoning traces) are emitted in real time for UI updates and monitoring.

\textbf{Customizing Agent Context with Skills and Prompts.}
\texttt{AgentContext} centralizes all inputs that shape LLM behavior, including prefixes/suffixes for system/user messages and user-defined \texttt{Skill} objects. 
Skills can be defined programmatically or loaded from markdown files (e.g., \texttt{.openhands/skills/}, or compatible formats like \texttt{.cursorrules}, \texttt{agents.md}). 
Each skill may always be active (\texttt{trigger=None}) to persistently augment the system prompt, or conditionally activated via keyword matching based on user input; skills may also include MCP tools. 
This design enables rich contextual and behavioral customization without modifying agent logic.

\textbf{Sub-Agent Delegation.}
The SDK supports hierarchical agent coordination through a delegation tool that demonstrates the extensibility of the tool abstraction. 
Sub-agents operate as independent conversations that inherit the parent’s model configuration and workspace context, enabling structured parallelism and isolation without any changes to the core SDK. 
The current implementation provides blocking parallel execution, implemented as a standard tool in the \texttt{openhands.tools} package, where the parent agent spawns and monitors sub-agents until all tasks complete. 
This pattern exemplifies how complex coordination behaviors—such as asynchronous delegation, dynamic scheduling, or fault-tolerant recovery—can be implemented entirely as user-defined tools, reinforcing the SDK’s design principle for extensibility that advanced agent orchestration requires no modification to the core framework.

\subsection{Context Window Management}
\label{sec:context}

To ensure the ever-growing history fits inside the LLM's context, the \texttt{Condenser} system drops events and replaces them with summaries whenever the history grows too large. 

The results of any given condensation are stored in the event log as a \texttt{CondensationEvent}. Before sending the event history to the LLM, the agent \emph{applies} these condensation events by removing forgotten events and inserting summaries. This strategy lets the SDK preserve the entirety of the event log, regardless of condensation, while also keeping the condenser implementations stateless.

Condensers enable long-running conversations and, as a mechanism for constraining the tokens consumed per step, can reduce the overall cost of a conversation. \texttt{LLMSummarizingCondenser} (the default condenser) has been shown to reduce API costs by up to $2 \times$ with no degradation in agent performance \citep{smith2025-openhands-context-condensensation}.

\subsection{Local Conversation}
\label{sec:local-conversation}

\texttt{LocalConversation} provides the simplest and most direct execution mode of the SDK, designed for rapid iteration and debuggability. It runs the full agent loop—LLM calls, tool invocation, event callbacks, and state updates—entirely in-process without network or container overhead. The core API offers intuitive control: developers can initialize a conversation with \texttt{Conversation(agent, workspace)}, send inputs using \texttt{send\_message()}, start execution via \texttt{run()}, pause and resume with \texttt{pause()} and \texttt{run()} respectively, inspect results through event callbacks or direct access of \texttt{.state} attribute.
Pausing automatically persists state and emits a \texttt{PauseEvent}, allowing agents to resume from the same point later.

\subsection{Secret Registry}
\label{sec:secret-registry}

\texttt{SecretRegistry} provides secure, late-bound, and remotely manageable credentials for tool execution. Each conversation maintains its own instance, ensuring strict per-session isolation. Tools access secrets only at execution time, and all secret values appearing in outputs are masked to prevent leakage. For example, the Bash Tool scans commands for secret keys, exports the referenced ones as environment variables, and replaces their occurrences in results with a constant mask (\texttt{<secret-hidden>}). Secrets may be static values or callables (e.g., token refreshers) and can originate from local stores or HTTP-based sources with secret-aware header handling. New custom sources can be easily added to integrate with external secret stores. All secrets are redacted during serialization and can be encrypted with a configurable cipher. They can also be updated mid-conversation—locally or through the agent server API—supporting live rotation without restarting agents.

\subsection{Security and Confirmation}
\label{sec:security-confirmation}

AI agents can perform risky actions, making security especially important when they run on a user’s machine. To address this, SDK treats security as a first-class concern within the agent’s control loop. Two abstractions form the core of this design: the \texttt{SecurityAnalyzer}, which rates each tool call as low, medium, high, or unknown risk, and the \texttt{ConfirmationPolicy}, which determines whether user approval is required before execution based on the action’s details and assessed risk.

When approval is required, the agent pauses in a special \texttt{WAITING\_FOR\_CONFIRMATION} state until the user explicitly approves or rejects the action. Upon rejection, it may retry with safer alternatives. The confirmation policy can be updated dynamically during a session, enabling adaptive trust—such as relaxing restrictions for safe, read-only operations like grep.

This architecture separates risk assessment from enforcement, allowing developers define custom \texttt{SecurityAnalyzer} and \texttt{ConfirmationPolicy} without touching tool executors or core logic. The SDK includes a built-in pair: \texttt{LLMSecurityAnalyzer}, which appends a \texttt{security\_risk} field to tool calls, and \texttt{ConfirmRisky} policy, which blocks actions exceeding a configurable risk threshold (default: high). 

\subsection{Deployment Architecture: Local to Remote}
\label{sec:deployment}

OpenHands Agent SDK's key innovation is seamless transition from local prototyping to remote production with minimal code changes. 
This is enabled by the agent API server and the abstraction of \texttt{Conversation} and \texttt{Workspace}.

\textbf{Conversation Factory: Local \& Remote Conversation.}
The \texttt{Conversation} class serves as a factory entry point that abstracts over local and remote execution. When instantiated with a string path or \texttt{LocalWorkspace}, it returns a \texttt{LocalConversation} that executes the full agent loop in-process, directly invoking tools and updating state synchronously (\sref{sec:local-conversation}).
When provided a \texttt{RemoteWorkspace}, the same call transparently constructs a \texttt{RemoteConversation}, which serializes the agent configuration and delegates execution to an agent server over HTTP and WebSocket. 
Both implementations share an identical API allowing seamless migration from local prototyping to containerized multi-user deployments without code changes. This factory pattern encapsulates all environment-specific logic behind a unified interface, achieving the SDK’s goal of “local-first, deploy-anywhere” development: the same agent code that runs interactively in a notebook can scale to distributed production simply by swapping the workspace type.

\begin{figure}[!h]
\centering
\begin{minted}[fontsize=\scriptsize, breaklines, linenos]{diff}
--- local.py
+++ remote.py
@@
 from openhands.sdk import LLM, Conversation
 from openhands.sdk.preset.default import get_default_agent
 llm = LLM(model="anthropic/claude-sonnet-4.1", ...)
 agent = get_default_agent(llm=llm)
-conversation = Conversation(agent=agent)
-conversation.send_message("Create hello.py")
-conversation.run()
+from openhands.workspace import DockerWorkspace
+with DockerWorkspace(...) as workspace:
+  conversation = Conversation(agent=agent, workspace=workspace)
+  conversation.send_message("Create hello.py")
+  conversation.run()
\end{minted}
\caption{Local-to-remote transition requires only importing and instantiating \texttt{DockerWorkspace}. All other code (agent configuration, LLM setup, message handling) remains unchanged.}
\label{fig:local-remote-code}
\end{figure}

\textbf{Agent Server.}
The \texttt{agent\_server} module implements an API server for remote agent execution (\fref{fig:agent-server-arch}). It exposes REST endpoints for conversation control (e.g., \texttt{POST /conversations}, \texttt{GET /conversations/{id}}) and WebSocket for event streaming. When a \texttt{RemoteConversation} starts, it serializes agent configuration—including LLM settings, tools, and context—into JSON and submits it to \texttt{/conversations}. The server reconstructs the agent, launches a local execution loop, and streams structured events back in real time, enabling responsive UIs without polling overhead.

\begin{figure}[!h]
\centering
\small
\resizebox{0.95\linewidth}{!}{%
\begin{tikzpicture}[
  node distance=0.5cm and 0.8cm, %
  every node/.style={font=\scriptsize, inner sep=2pt},
  box/.style={rectangle, draw, thick, minimum width=1.4cm, minimum height=0.6cm, align=center},
  client/.style={box, fill=blue!10},
  server/.style={box, fill=purple!10},
  internal/.style={box, fill=green!10},
  arrow/.style={->, thick, >=Latex, shorten >=1pt}
]

\node[client] (client) {RemoteConversation\\(Client)};

\node[server,  right=1.2cm of client] (server) {Agent\\Server};
\node[internal,right=0.6cm of server] (local)  {Local\\Conversation};
\node[internal,right=0.6cm of local]  (agent)  {Agent +\\Tools};

\draw[dashed, thick, rounded corners=2pt]
  ([xshift=-0.1cm, yshift=0.3cm]server.north west)
  rectangle
  ([xshift=0.1cm,  yshift=-0.3cm]agent.south east);
\node[below=0.01cm of local, font=\tiny] {Container};

\draw[arrow] (client) -- node[midway, above=3mm, font=\tiny]{HTTP/WS} (server);
\draw[arrow] (server) -- node[midway, above=3mm, font=\tiny]{spawns} (local);
\draw[arrow] (local)  -- node[midway, above=3mm, font=\tiny]{executes}  (agent);

\end{tikzpicture}%
}
\caption{Agent server architecture. Client serializes agent configuration via HTTP; the server executes using SDK components inside the container and streams events via WebSocket.}
\label{fig:agent-server-arch}
\end{figure}

To support scalable and isolated execution, we provide official Docker images that bundle the full agent-server stack—including the API server, VSCode Web, VNC desktop, and Chromium browser. Each agent instance runs in an independent container with a dedicated file system, environment, and resource. This containerized design simplifies deployment and enables SaaS-style multi-tenancy while preserving workspace isolation.

\begin{figure}[!h]
\begin{minted}[fontsize=\scriptsize, breaklines, linenos]{python}
class BaseWorkspace(ABC):
    def __enter__(self) -> "BaseWorkspace"
    def __exit__(self, exc_type: Any, exc_val: Any, exc_tb: Any) -> None
    def execute_command(self, command: str) -> CommandOutput:
        """Run shell command in workspace environment."""
    def file_upload(self, path: str, content: bytes) -> None:
        """Write file to workspace filesystem."""
    def file_download(self, path: str) -> bytes:
        """Read file from workspace filesystem."""

class LocalWorkspace(BaseWorkspace):
    def execute_command(self, cmd: str) -> CommandOutput:
        return subprocess.run(cmd, shell=True, capture_output=True)

class RemoteWorkspace(BaseWorkspace):
    def execute_command(self, cmd: str) -> CommandOutput:
        return self._api_client.post("/execute", json={"command": cmd})

# Workspace factor class
class Workspace:
    def __new__(cls, *, host=None, working_dir="workspace/project", api_key=None):
        if host is provided:
            return RemoteWorkspace(working_dir, host, api_key)
        else:
            return LocalWorkspace(working_dir)
\end{minted}
\caption{Workspace interface. Implementations handle environment details.}
\label{fig:workspace-interface}
\end{figure}

\textbf{Workspace.}
\label{sec:workspace}
The \texttt{BaseWorkspace} abstract class enables sandboxes for agents.
\textbf{Local Workspace} executes in-process against the host filesystem and shell; it is effectively a thin, no-op wrapper that forwards file/command/git operations directly, enabling fast prototyping without network hops. 
\textbf{Remote Workspace} preserves the same interface but delegates all operations over HTTP to an Agent Server (see \fref{fig:workspace-interface}), with concrete sponsors including a containerized server (\texttt{DockerWorkspace}) or a API-managed runtime (\texttt{APIRemoteWorkspace}). The factory \texttt{Workspace(\ldots)} resolves to local when only \texttt{working\_dir} is provided and to remote when \texttt{host}/runtime parameters are present, ensuring the agent code remains unchanged across environments.

\section{Reliability and Evaluation}
\label{sec:evaluation}

We assess the OpenHands Agent SDK through four complementary lenses: (1) production reliability comparing V0 and V1, (2) systems overhead of the event-sourcing mechanism, (3) continuous testing, and (4) benchmark evaluation across diverse models and tasks.

\subsection{Production Reliability: V0 vs.\ V1}
\label{sec:reliability}

To directly evaluate whether the architectural redesign improves operational reliability, we analyzed production logs from a 15-day parallel rollout in which V0 and V1 served users simultaneously. Exceptions were extracted from conversation failure logs and categorized into three classes (\tref{tab:error-taxonomy}):

To ensure comparability, we report \emph{system-attributable errors}---error classes directly attributable to agent infrastructure or SDK logic, excluding LLM provider errors that are external to both architectures. Results are shown in \tref{tab:reliability-results}.

\begin{table}[!h]
\centering
\small
\caption{System-attributable errors per 1k conversations during 15-day production rollout.}
\label{tab:reliability-results}
\begin{tabular}{@{}l r r@{}}
\toprule
& \textbf{V0 (Legacy)} & \textbf{V1 (SDK)} \\
\midrule
Infrastructure errors & 69.8\,/\,1k & 0.0\,/\,1k \\
SDK errors & N/A & 29.7\,/\,1k \\
\midrule
System-attributable error rate & 78.0\,/\,1k & 30.0\,/\,1k \\
\bottomrule
\end{tabular}
\end{table}

\textbf{Result:} V1 reduces system-attributable failures by 61\% (from 78.0 to 30.0 errors per 1k conversations).
The eliminated V0 infrastructure errors were dominated by: \texttt{HTTPStatusError~(401)} at 43.0\,/\,1k (inter-pod authentication failures), \texttt{AgentRuntimeNotReadyError} at 18.8\,/\,1k (runtime pod readiness races), and connection/timeout errors at 3.1\,/\,1k (network instability).
These failures stem from V0's inter-pod HTTP communication between the conversation manager and execution runtime. V1's co-located execution model eliminates these observed failure modes by removing this dependency entirely.
The remaining V1 SDK errors (29.7\,/\,1k) were dominated by a condensation bug discovered during the rollout of extended thinking support, as the LLM provider introduced additional constraints on the event/message interface. This issue has since been fixed in the official release.

\subsection{Event-Sourcing Systems Overhead}
\label{sec:event-sourcing-overhead}

To quantify the I/O cost of event sourcing, we replayed real payloads from 433 SWE-Bench Verified conversations (39,870 events) through the production \texttt{LocalFileStore} path (\tref{tab:event-sourcing-overhead}). All persist and recovery latencies are negligible relative to LLM round-trip times (typically 1--30\,s), storage grows linearly, and crash recovery completes in under 20\,ms even for the longest observed conversation (358 events). These results demonstrate that event sourcing adds minimal overhead to the agent execution loop.

\begin{table}[!h]
\centering
\caption{Event-sourcing systems overhead measured on SWE-Bench Verified traces (433 conversations, 39,870 events).}
\label{tab:event-sourcing-overhead}
\resizebox{\columnwidth}{!}{%
\begin{tabular}{@{}lrrr@{}}
\toprule
\textbf{Metric} & \textbf{Median} & \textbf{P95} & \textbf{At max (358 events)} \\
\midrule
Per-event persist latency & 0.20\,ms & 0.31\,ms & --- \\
Action cycle persist (Action+Obs) & 0.40\,ms & 0.56\,ms & --- \\
Full state replay & 4.1\,ms & 9.7\,ms & 18.9\,ms \\
Crash recovery (replay + unmatched-action scan) & 7.4\,ms & 14.9\,ms & 32.1\,ms \\
Storage per conversation & 380\,KB & 1.4\,MB & 3.4\,MB \\
\bottomrule
\end{tabular}%
}
\end{table}

\subsection{Continuous Quality Assurance}

The SDK employs a three-tier testing strategy that balances coverage, cost, and depth:
\begin{itemize}[itemsep=2pt,topsep=2pt,parsep=0pt,partopsep=0pt,leftmargin=8pt]
  \item \textbf{Programmatic Tests} — Run on every commit. These tests mock llm calls and verify core logic, data flow, and API contracts within seconds. Mocking allows quicker feedback, ensuring that most regressions are caught before any external API calls are made.
  \item \textbf{LLM-based Tests} — Include both integration and example tests (see below). Executed daily and on-demand for pull requests. These tests use real models (Claude~Sonnet~4.5, GPT-5~Mini, DeepSeek~Chat) to validate reasoning, tool invocation, and environment stability. Each run costs \$0.5–\$3 and completes in less than 5 minutes.
  \item \textbf{Benchmark Evaluation} — On-demand, high-cost evaluations (\$100--1000, hours per run) that measure comprehensive agent capabilities on academic datasets.
\end{itemize}

\textbf{Integration tests} cover multiple scenario-based workflows (e.g., file manipulation, command execution, git operations, and browsing), while \textbf{example tests} periodically run all SDK examples (custom tools, MCP integration, persistence, async execution, routing, etc) to ensure end-to-end reliability.
The suite is continuously expanded as new agent behaviors and failure patterns are discovered, improving coverage and regression sensitivity over time. 
On-demand execution for these LLM-based tests further optimizes CI/CD cost: integration tests target high-risk changes, example tests cover user-facing modules, and daily runs track regressions across overall codebase updates.

\begin{figure}[!ht]
\begin{minted}[fontsize=\scriptsize, breaklines, linenos]{python}
class BaseIntegrationTest(ABC):
    def setup(self): """Prepare test environment"""
    def tools(self) -> list[Tool]: """Agent capabilities"""
    
    def verify_result(self) -> bool: """Check success"""
    def run(self, instruction: str) -> bool:
        self.setup()
        conversation = Conversation(Agent(llm, self.tools), workspace)
        conversation.send_message(instruction)
        conversation.run()
        return self.verify_result()
\end{minted}
\vspace{-18pt}
\caption{Integration test framework. Subclasses implement \texttt{setup()}, \texttt{tools}, and \texttt{verify\_result()} for specific scenarios.}
\label{fig:base-integration-test}
\end{figure}

\subsection{Benchmark Evaluation}

We validate the SDK on academic benchmarks along two axes: first, that the V0-to-V1 architectural transition preserves (and, where the architecture enables new provider features, improves) agentic capability; second, that the SDK generalizes across models and task types.

The evaluation spans five categories of software engineering work: Issue Resolution (SWE-Bench Verified~\citep{jimenez2024swebench}), Greenfield Development (Commit0~\citep{commit0}), Frontend Development (SWE-Bench Multimodal~\citep{yang2024swebenchmultimodal}), Software Testing (SWT-Bench~\citep{mundler2024swtbench}), and Information Gathering (GAIA~\citep{mialon2023gaiabenchmarkgeneralai}). Results for all 14 evaluated models are maintained on the continuously updated OpenHands Index.\footnote{\url{https://index.openhands.dev}}

\textbf{V0 vs.\ V1: Capability Preservation.}
We compare V0 and V1 on SWE-Bench Verified using matched models to isolate the contribution of the SDK redesign (\tref{tab:v0-v1-benchmarks}). With Claude Sonnet~4, V0 and V1 achieve identical performance ($68.0\%$), confirming that the architectural redesign preserves baseline agentic capability. With Sonnet~4.5, V1 gains $+8.2$ points; we attribute this to extended thinking support, which V1's event-sourced architecture integrates naturally but would require significant engineering effort to retrofit into V0's multi-component design.

\begin{table}[h]
\centering
\small
\begin{tabular}{@{}l c c@{}}
\toprule
\textbf{Model} & \textbf{V0} & \textbf{V1 (SDK)} \\
\midrule
Claude Sonnet 4.5 & $64.6\%$ & $\mathbf{72.8\%}$ \\
Claude Sonnet 4 & $68.0\%$ & $68.0\%$ \\
\bottomrule
\end{tabular}
\caption{V0 vs.\ V1 on SWE-Bench Verified with matched models. Sonnet~4 confirms capability parity; the Sonnet~4.5 gain is attributed to extended thinking support enabled by V1's architecture.}
\label{tab:v0-v1-benchmarks}
\end{table}

\textbf{Comprehensive Multi-Model Evaluation.}
We evaluated 14 language models---7 closed-source (Anthropic, OpenAI, Google) and 7 open-weights (MiniMax, DeepSeek, Zhipu AI, Moonshot AI, Alibaba, NVIDIA)---across all five categories. \tref{tab:multi-model-eval} summarizes the best SDK result per category alongside published state-of-the-art. The SDK achieves SOTA on 3 of 5 benchmarks and remains competitive on the other two (within $2.6$ and $5.2$ points), using a single model per evaluation while several SOTA systems employ multi-model orchestration.

\begin{table}[h]
\centering
\footnotesize
\resizebox{\columnwidth}{!}{%
\begin{tabular}{@{}l c c c c c@{}}
\toprule
& \textbf{Issue Resol.} & \textbf{Greenfield} & \textbf{Frontend} & \textbf{Testing} & \textbf{Info.\ Gather.} \\
& \footnotesize SWE-Bench V. & \footnotesize Commit0 & \footnotesize SWE-Bench MM & \footnotesize SWT-Bench V. & \footnotesize GAIA (test) \\
\midrule
Published SOTA & $79.2\%$ & $12.5\%$ & -\footnotemark & $84.0\%$ & $74.6\%$ \\
\midrule
Best SDK & $76.6\%$ & $\mathbf{56.2\%}$ & $\mathbf{44.1\%}$ & $78.8\%$ & $\mathbf{80.0\%}$ \\
\textit{Model} & \textit{Opus 4.5} &  \textit{GPT-5.4} & \textit{Gemini 3.1 Pro} &  \textit{Opus 4.6} &  \textit{Opus 4.6} \\
\midrule
2nd SDK & $75.6\%$ & $56.2\%$ & $41.8\%$ & $78.5\%$ & $78.8\%$ \\
 \textit{Model} &  \textit{GPT-5.4} &  \textit{Opus 4.6} &  \textit{Opus 4.6} &  \textit{Opus 4.5} &  \textit{GPT-5.4} \\
\bottomrule
\end{tabular}
}
\caption{Multi-model benchmark evaluation across five SE task categories (14 models evaluated). Bold indicates the SDK exceeds published SOTA. Full per-model results are available on the OpenHands Index (\url{https://index.openhands.dev}).}
\label{tab:multi-model-eval}
\end{table}
\footnotetext{OpenHands Index uses the test dataset of SWE-bench Multimodal, for which no state-of-the-art results have yet been published.}

\textbf{SDK-Enabled Insights.}
The uniform evaluation harness enables observations that would be difficult without a model-agnostic SDK. Models exhibit clear task-specific specialization: Claude models dominate issue resolution and testing, while GPT-5.4 leads in long-horizon greenfield development ($62.5\%$, $+12.5$ pts ahead of second place).

\begin{table*}[!th]
\centering
\resizebox{\textwidth}{!}{%
\begin{tabular}{@{}l|c|c|c|c|c@{}}
\toprule
& \makecell{\textbf{OpenAI}\\\textbf{Agents SDK}}
& \makecell{\textbf{Claude}\\\textbf{Agent SDK}}
& \makecell{\textbf{Google Agent}\\\textbf{Dev.\ Kit}}
& \makecell{\textbf{LangChain/}\\\textbf{LangGraph}}
& \makecell{\textbf{OpenHands}\\\textbf{Agent SDK}} \\
\midrule
\multicolumn{6}{@{}l}{\textit{\textbf{Standalone SDK Features}}} \\
MCP Support & \cmark & \cmark & \cmark & \cmark & \cmark \\
Custom Tools & \cmark & \cmark & \cmark & \cmark & \cmark \\
History Persistence \& Restore & \cmark & \cmark & \cmark & \cmark & \cmark \\
Sub-agent Delegation & \cmark & \cmark & \cmark & \cmark & \cmark \\
Model Agnostic (100+ LLMs) & \cmark & \xmark & \cmark & \cmark & \cmark \\
Multi-LLM Routing & \xmark & \xmark & \xmark & \simmark & \cmark \\
Conversation Cost \& Token Tracking & \cmark & \cmark & \xmark & \cmark & \cmark \\
Pause/Resume Agent Execution & \cmark & \xmark & \xmark & \cmark & \cmark \\
Native Support for Non-function-calling models & \xmark & \xmark & \xmark & \xmark & \cmark \\
Security Analyzer for Agent Action & \xmark & \xmark & \cmark & \xmark & \cmark \\
Action Confirmation Policies & \simmark & \xmark & \simmark & \simmark & \cmark \\
Context Files (e.g., repo.md, AGENTS.md) & \xmark & \cmark & \xmark & \xmark & \cmark \\
Agent Skills  & \xmark & \cmark & \xmark & \xmark & \cmark \\
Context Condensation & \xmark & \cmark & \xmark & \cmark & \cmark \\
Secrets Management with Auto-Masking & \xmark & \xmark & \xmark & \xmark & \cmark \\
Agent Stuck Detection & \xmark & \xmark & \xmark & \xmark & \cmark \\
\midrule
\multicolumn{6}{@{}l}{\textit{\textbf{Production Server \& Deployment}}} \\
Builtin REST+WebSocket Server & \xmark & \xmark & \xmark & \cmark & \cmark \\
Session-Based Authentication & \xmark & \xmark & \xmark & \cmark & \cmark \\
Builtin Remote Agent Execution & \xmark & \xmark & \xmark & \cmark & \cmark \\
Agent Environment Sandboxing & \xmark & \xmark & \simmark & \xmark & \cmark \\
\midrule
\multicolumn{6}{@{}l}{\textit{\textbf{Continuous Quality Assurance}}} \\
Programmatic tests & \cmark & \cmark & \cmark & \cmark & \cmark \\
Strong Type Checking & \cmark & \cmark & \cmark & \cmark & \cmark \\
Frequent LLM-based tests & \xmark  & \cmark & \cmark & \xmark & \cmark \\
Built-in Academic Benchmark Evaluation & \xmark & \xmark & \xmark & \xmark & \cmark \\
\bottomrule
\end{tabular}%
}
\caption{
Feature comparison across major agent SDKs (\cmark = full support, \simmark = partial, \xmark = absent).
\textbf{Methodology:} Features were assessed from official documentation, public repositories, and release notes as of October 2025.
Versions assessed: OpenAI Agents SDK v0.4.2, Claude Agent SDK v0.1.6, Google ADK v1.17.0, LangChain v1.0.3 / LangGraph v1.0.2, OpenHands Agent SDK v1.0.0.
``Partial'' indicates the capability requires significant external setup or community extensions rather than built-in support.
OpenHands uniquely combines: \textbf{(i)} native remote execution with environment sandboxing,
\textbf{(ii)} LLM-powered action-level security analysis, 
\textbf{(iii)} model-agnostic multi-LLM routing with first-class support for non-function-calling models, and
\textbf{(iv)} built-in academic benchmark evaluation for continuous quality assurance.
}
\label{tab:sdk-comparison}
\end{table*}

\section{Related Work}
\label{sec:sdk-comparison}

Recent agent SDKs share common goals of tool orchestration, state management, and production deployment. 
General-purpose frameworks such as LangChain and LangGraph focus on compositional pipelines and stateful graph execution with durable checkpoints for long-running reasoning workflows \citep{langchain,langgraph}. LangGraph Platform further offers deployment and server capabilities, making it one of the most feature-complete open-source alternatives.
Provider SDKs from OpenAI and Anthropic emphasize production orchestration, guardrails, and handoffs within their own model ecosystems \citep{openai-agents,anthropic-agents}, while Google's Agent Development Kit (ADK) targets model-agnostic agents integrated with Vertex AI for managed deployment \citep{google-adk}. 

In contrast, the \emph{OpenHands Software Agent SDK} is a fully open-source, vendor-agnostic platform purpose-built for \emph{software engineering agents}. 
It couples an event-sourced state model and immutable configuration with sandboxed execution, a built-in REST/WebSocket server, and workspace-level remote interfaces for human-agent collaboration. 
These design choices enable reproducible, deterministic execution and seamless transition from local prototyping to production-scale deployments. 
A feature-level comparison with major agent SDKs---including LangChain/LangGraph---is provided in \tref{tab:sdk-comparison}, with comparison methodology detailed in the caption.

\section{Discussion}
\label{sec:discussion}

\paragraph{Architectural Alternatives Considered.}
During the V1 redesign, we evaluated several alternative approaches. For state management, we considered a traditional database-backed model (used by many web frameworks) but rejected it because it would couple the SDK to a specific storage backend and make offline replay difficult; event sourcing was chosen for its reproducibility and storage-agnostic design. For execution isolation, we considered mandatory containerization (as in V0) and fully local-only execution; the optional isolation model was selected as a compromise that aligns with MCP's local-first assumptions while preserving sandboxing for production. For the package structure, we considered a monolithic SDK (simpler dependency management) versus a microservices architecture (maximum flexibility); the four-package design balances independent testability with practical dependency management.

\paragraph{Multi-Tenant Deployment.}
The current security and isolation discussion focuses primarily on single-user scenarios. In multi-tenant deployments---where multiple users submit requests to the same agentic application or multiple applications share infrastructure components---additional considerations apply. The containerized deployment model (\sref{sec:deployment}) provides process-level isolation between tenants, with each agent instance running in an independent container with a dedicated filesystem and resource limits. The event-sourced state model naturally supports tenant isolation since each conversation maintains its own event log. However, shared resources such as LLM API keys, MCP servers, and secret registries require careful access control. The SDK's \texttt{SecretRegistry} supports per-conversation secrets, and the agent server's session-based authentication provides a foundation for multi-tenant access control, but a comprehensive multi-tenant security audit remains future work.

\paragraph{Limitations.}
The current implementation focuses on single-agent conversations. While the event-sourced architecture naturally supports interleaving events from multiple agents, coordination mechanisms for multi-agent collaboration require further design. The security framework, while substantially improved over V0, cannot guarantee complete safety: LLM-based security analysis is subject to adversarial prompts and inconsistent classification.

\section{Conclusion}
The OpenHands Software Agent SDK provides a reference architecture for production agent systems, bridging the gap between rapid local prototyping and production deployment through a stateless, event-sourced, and composable design spanning four packages (SDK, Tools, Workspace, Server). 
From our experience scaling OpenHands from research prototype to production system, we distill four key lessons: strict separation between core agent logic and applications is essential for maintainability; event-sourced state enables reproducibility and fault recovery with negligible overhead (sub-millisecond persist latency, crash recovery under 20\,ms); immutable component design prevents configuration drift; and a unified execution model supporting both local and sandboxed deployment streamlines the transition from experimentation to production.
These architectural choices yield measurable reliability gains: a 15-day production comparison demonstrates a 61\% reduction in system-attributable failures relative to V0, while evaluations across 14 models and five benchmark categories confirm that the redesign preserves and enables strong agentic performance, achieving state-of-the-art results on three of five tasks. Together, these results validate the SDK as a robust foundation for both research and industrial-scale deployment.

\clearpage
\newpage
\bibliographystyle{plainnat}
\bibliography{custom}

\newpage
\beginappendix

\begin{table}[!h]
\centering
\small
\caption{Error taxonomy for production reliability analysis.}
\label{tab:error-taxonomy}
\begin{tabular}{@{}l p{5.5cm}@{}}
\toprule
\textbf{Category} & \textbf{Definition \& Examples} \\
\midrule
Infrastructure & Runtime communication and orchestration failures (e.g., \texttt{HTTPStatusError}, \texttt{AgentRuntimeNotReadyError}, \texttt{ConnectionError}) \\
SDK & Internal SDK logic errors (e.g., \texttt{CondensationError}) \\
LLM Provider & External API failures, analyzed separately (e.g., \texttt{RateLimitError}, \texttt{AuthenticationError}) \\
\bottomrule
\end{tabular}
\end{table}

\end{document}